\documentclass[prl,notitlepage,twocolumn]{revtex4}

\usepackage{amsmath}
\usepackage{amssymb}
\usepackage{bm}
\usepackage{graphicx}
\usepackage{times}

\begin{document}

\title{Strong coupling between excitons in transition metal dichalcogenides and optical bound states in the continuum}

\author{K.\,L. Koshelev$^{1,2}$}
\email{ki.koshelev@gmail.com}
\author{S.\,K. Sychev$^{1}$}
\author{Z.\,F. Sadrieva$^{1}$}
\author{A.\,A. Bogdanov$^{1}$}
\author{I.\,V. Iorsh$^{1}$}

\affiliation{$^{1}$ITMO University, 197101 St.~Petersburg, Russian Federation}
\affiliation{$^{2}$Nonlinear Physics Centre, Australian National University, Canberra ACT, Australia}
\date{\today}

\begin{abstract}
Being motivated by recent achievements in the rapidly developing fields of  optical bound states in the continuum (BICs) and excitons in monolayers of transition metal dichalcogenides, we analyze strong coupling between BICs in $\rm Ta_2O_5$ periodic photonic structures and excitons in $\rm WSe_2$ monolayers. We demonstrate that giant radiative lifetime of BICs allow to engineer the exciton-polariton lifetime enhancing it three orders of magnitude compared to a bare exciton. We show that maximal lifetime of hybrid light-matter state can be achieved at any point of $\mathbf{k}$-space by shaping the geometry of the photonic structure. Our findings open new route for the realization of the moving exciton-polariton condensates with non-resonant pump and without the Bragg mirrors which is of paramount importance for polaritonic devices.
\end{abstract}
\keywords{}

\maketitle

Monolayers of transition metal dichalcogenides (TMDCs) are a certain class of post-graphene two-dimensional materials~\cite{Bhimanapati2015}, attracting vast research interest in recent years. TMDC are direct-gap semiconductors, exhibiting strong light-matter coupling~\cite{Mak2016}. Moreover, these structures support excitons characterized by both large binding energies and sufficiently large Bohr radii~\cite{Wang2017}. While the former leads to the existence of strong excitonic response at room temperature, the latter provides strong optical nonlinearity due to the exciton-exciton interactions~\cite{Shahnazaryan2017}. Another important property of the TMDC excitons is the large oscillator strength leading to the substantial  exciton-photon interaction in these structures. These properties allow the observation of the so-called strong coupling regime, leading to the emergence of the new quasiparticles, exciton-polaritons~\cite{Kavokin2017} at room temperatures in structures comprising TMDC monolayer and an optical cavity. 

Excitons-polaritons have been extensively studied in last two decades both due to their fascinating fundamental properties, such as high-temperature Bose condensation and superfluidity~\cite{Kasprzak2006,Amo2009} as well as well as emerging applications such as virtually thresholdless polariton lasers~\cite{Bhattacharya2014} and energy effective all optical logic gates~\cite{Amo2010}. Strong coupling of TMDC excitons to light has been observed in the structures resembling the conventional microcavities, where the monolayer was sandwiched between two Bragg mirrors~\cite{Liu2015,Dufferwiel2015,Sun2017,Dufferwiel2017}. At the same time, since fabrication of high quality TMDC monolayers is based on  the mechanical exfoliation techniques, and thus not compatible with the standard epitaxial techniques used for the Bragg mirror fabrication, the realization of the structures considered in~\cite{Liu2015,Dufferwiel2015,Sun2017,Dufferwiel2017} is quite technologically demanding. It would be thus extremely useful to realize high quality optical resonances without the requirement for the growth of the upper mirror. As an alternative, a whispering gallery mode (WGM) in a disc resonator could be used for the realization of the strong  coupling~\cite{Ye2015}. However, this approach also requires sophisticated fabrication techniques and the precise positioning of the monolayer over the region of the disc resonator where the WGM mode has antinodes. 

\begin{figure} [t]
\begin{minipage}{0.99\linewidth}
\begin{center}
\includegraphics[width=0.95\linewidth]{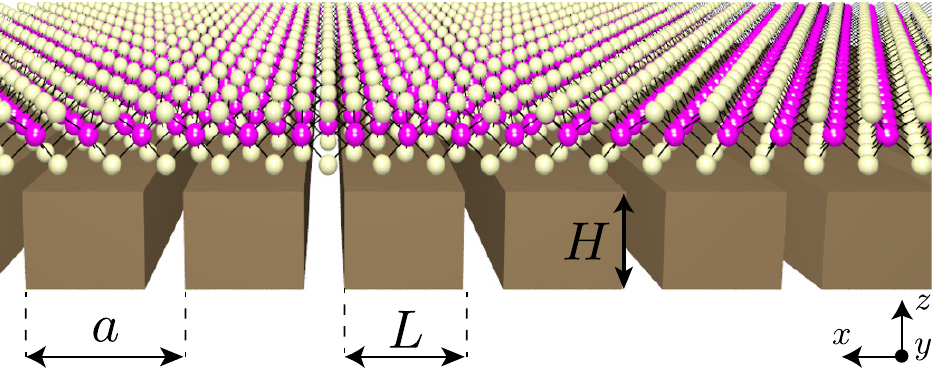}
\end{center}
\caption{Isometric view of a PCS-TMDC structure. The refractive index of $\rm Ta_2O_5$ bars is equal to $2.1$. The TMDC made of $\rm WSe_2$ is laid on top of the PhC slab. } 
\label{fig:1}
\end{minipage}
\end{figure} 

In this Letter we propose an alternative scheme for the realization of strong exciton-photon coupling in 2D materials without using mirrors which is beneficial both in terms of ease of realization and tunability. We focus on structures comprising of a TMDC and a photonic crystal slab (PCS). The idea behind that is the exploitation of the so-called optical bound state in the continuum (BIC) ~\cite{Hsu2016}, supported by the  PCS, as a high quality cavity mode. BICs in the periodic photonic structures originate due to destructive interference of the leaky modes supported by the PCS. They are characterized by the Bloch vector lying within the light cone in vacuum, high quality factors (up to 10$^6$ experimentally measured in photonic crystal slabs~\cite{Hsu2013}).  The bound states in the continuum in photonic crystal slabs and plasmonic lattices being high quality factor resonant modes have already found its applications for sensing~\cite{Yanik2011}, filtering~\cite{Foley2014} and lasing~\cite{Hirose2014,Kodigala2017}.

\begin{figure*} [t]
\begin{minipage}{0.99\linewidth}
\begin{center}
\includegraphics[width=0.95\linewidth]{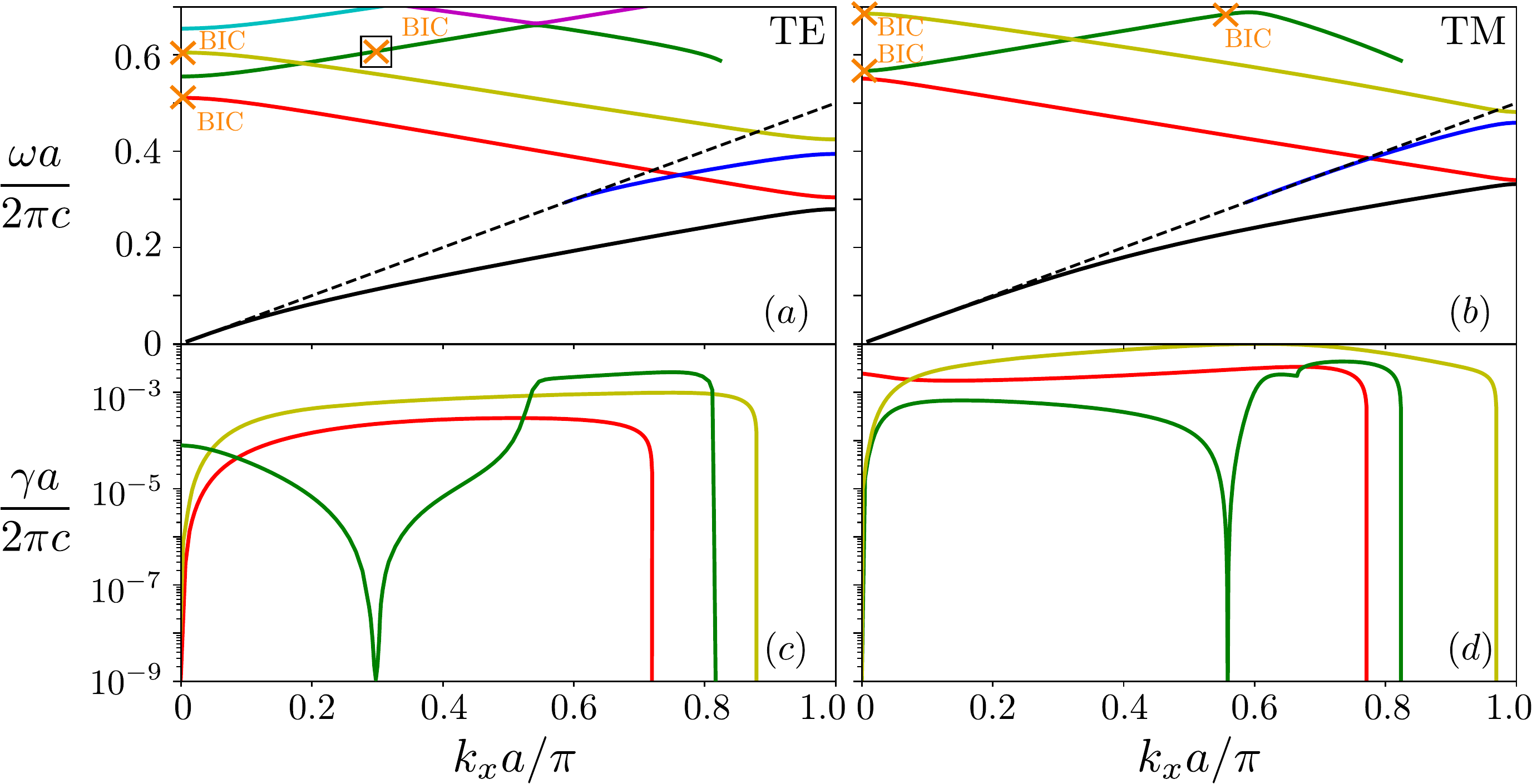}
\end{center}
\caption{ Eigenmode spectrum of air-suspended ${\rm Ta}_2{\rm O}_5$ one-dimensional grating. Band structure $\omega a / 2\pi c$ for (a) TE-polarized and (b) TM-polarized modes, respectively. Black dashed lines represent the light lines $\omega = c k_x$. Dimensionless inverse radiation lifetime  $\gamma a / 2\pi c$  for (c) TE-polarized and (d) TM-polarized modes, respectively.  BICs are marked by orange crosses.} 
\label{fig:2}
\end{minipage}
\end{figure*}

The design of structure is shown in Fig.~\ref{fig:1} - a $\rm WSe_2$ flake is placed on top of an one-dimensional $\rm Ta_2O_5$ PCS. We notice that strong light-matter coupling demands both high-Q photonic structures and long-living excitonic states. We address this problem by tuning the PCS shape and material parameters to the regime of  bound states in the continuum providing a giant quality factor of resonator which is limited by surface roughness and finite size of the sample only. 

We begin with analysis of the eigenmode spectrum of the PCS applying the guided-mode expansion (GME) method~\cite{andreani2006photonic} widely used for characterization of photonic eigenmodes of periodic photonic structures~\cite{galli2010low,minkov2017photonic,vasco2017statistics}. Within this approach the Maxwell's equations are treated by expanding the magnetic field into the basis of guided modes of an effective homogeneous waveguide resulting in a Hermitian eigenvalue equation which is solved numerically. Radiative losses of leaky modes with frequencies above the light line are calculated by coupling coefficients to the continuum of scattering modes of the effective waveguide using the perturbation theory and the Fermi's golden rule.

We consider an air-suspended ${\rm Ta}_2{\rm O}_5$ one-dimensional grating with lower and upper air claddings being semi-infinite. The PCS consists of rectangular bars with height $H$ and width $L$ being spaced equidistant with a period of $a$ (see Fig.~\ref{fig:1}). We put refractive index of ${\rm Ta}_2{\rm O}_5$ equal to 2.1 which is appropriate for the red band of the visible spectrum range. The calculations are performed for the PCS with $a = 1.03H$, $L=0.90a$ and the eigenvalue problem is truncated by 101 plane waves and 8 guided modes of the effective waveguide kept in the expansion~\cite{andreani2006photonic}.

The spectrum of eigenfrequencies $\omega$ and inverse radiation lifetimes $\gamma=1/(2\tau_{\rm rad})$ of the PCS for in-plane wavevectors along the $x$ direction of the first Brillouin zone is shown in Fig.~\ref{fig:2}(a,c) for TE-polarized and in Fig.~\ref{fig:2}(b,d) for TM-polarized modes, respectively. Dispersion curves under the light line $\omega=c k_x$ describe pure guided modes with zero diffraction losses while photonic states above the light line are leaky provided their radiation lifetime is finite. The BICs represent unusual leaky modes with $\gamma = 0$ and can be formed both at the center of the Brillouin zone (at-$\Gamma$ BIC) and at specific points between the zone edge and center  (off-$\Gamma$ BIC). Importantly, all types of BICs are topologically stable against perturbations of the PCS geometry~\cite{zhen2014topological}. Therefore, while we do not account for the material dispersion and the PCS band structure scales linearly as its size parameters change, the BIC position can be tuned, for example, by variation of structure height $H$.

Optical properties of $\rm WSe_2$ monolayers are governed by very robust excitons with binding energies of the order of 500 meV~\cite{he2014tightly}. We study A-type excitons representing bound states of electrons in the conduction band and holes in the upper subband of the valence band of $\mathbf{K}_{+}$ and $\mathbf{K}_{-}$ valleys~\cite{glazov2015spin}. Optical selection rules result in distinction of bright and dark excitons being active for in-plane and out-of-plane polarization of incident light, respectively. We focus on bright excitonic states, which represent, in general, a pair of valley-degenerate states with $\sigma_{\pm}$ polarization. In this case a linearly polarized pump excites a superposition of excitons with total polarization along the in-plane component of the electric field of light.  

\begin{figure} [t]
\begin{minipage}{0.99\linewidth}
\begin{center}
\includegraphics[width=0.95\linewidth]{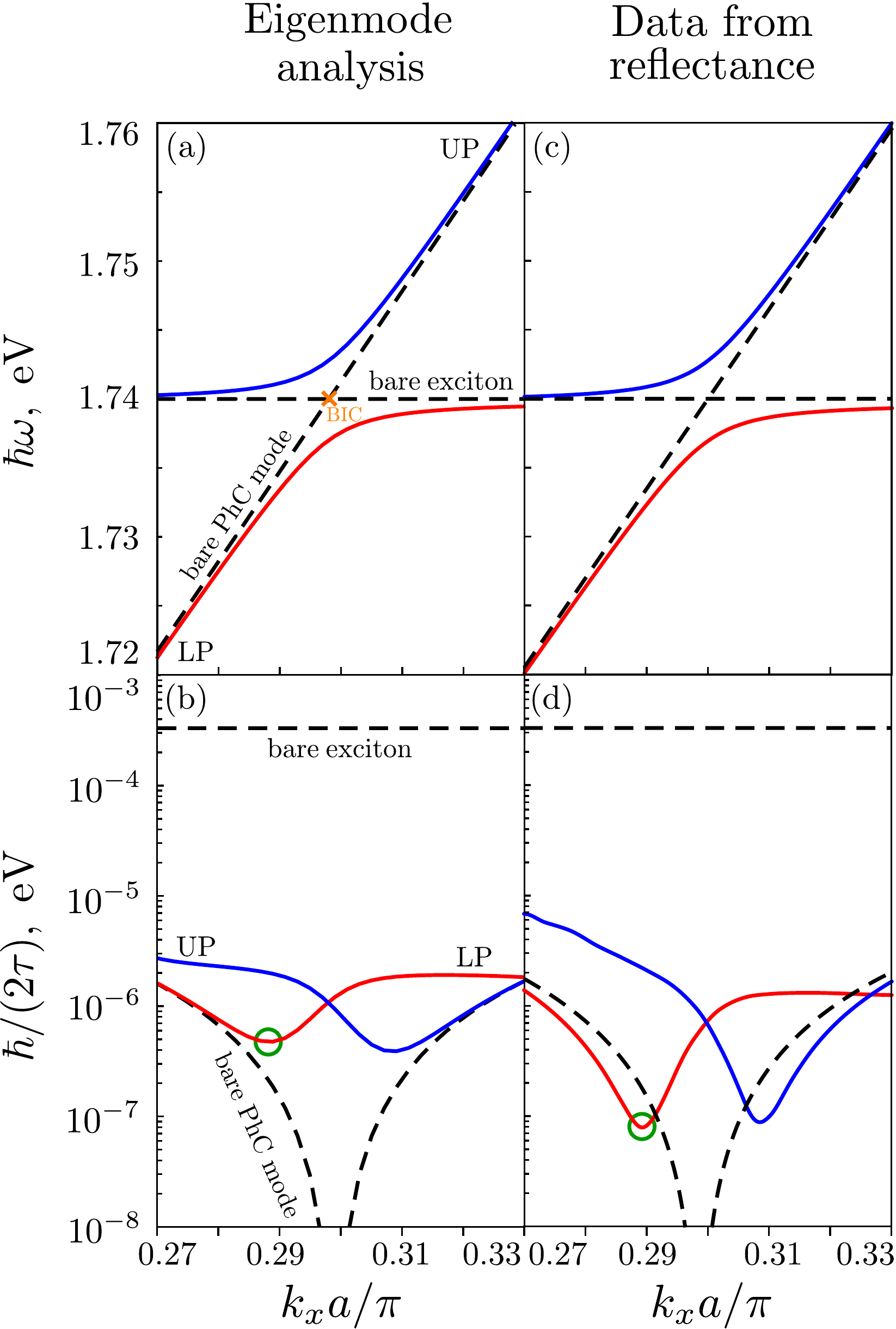}
\end{center}
\caption{Dispersion and inverse lifetime of exciton-polaritons at the conditions of strong coupling between the TE-polarized photonic mode supporting an off-$\Gamma$ BIC and in-plane polarized exciton with energy of $1.74$ eV calculated by using the GME method (a,b) and the fitting of Fano lineshape of reflection spectrum (c,d). Upper (UP) and lower (LP) polariton branches are shown with blue and red solid lines, respectively. Bare excitonic and photonic dispersion is depicted with black dashed lines. State of LP with maximal lifetime is marked by a green circle.}
\label{fig:3}
\end{minipage}
\end{figure} 

To investigate light-matter interaction between excitons in the TMDC monolayer and BICs in the PCS we apply the full quantum formalism for both photonic and excitonic states being an extension of the GME method \cite{andreani2007light, gerace2007quantum}. We start by the second quantization of both the electromagnetic field in a non-uniform dielectric environment and the exciton field which exhibits properties of a gas of non-interacting bosons while the density of excitons is much smaller than the saturation density. Coupling between photonic and exciton modes is governed by the oscillator strength $f$ and the overlap integral of mode profiles   
\begin{equation}
V_{n, \mathbf{k}_{\|}} = - i \sqrt{f}  \int\limits_{\rm cell}d\mathbf{r}_{\|}\   \mathbf{e} \cdot \mathbf{E}^{\rm up}_{n,\mathbf{k}_{\|}}(\mathbf{r}_{\|})e^{-i\mathbf{k_{\rm exc}}\mathbf{r}_{\|}}.
\label{eq:1}
\end{equation}
Here $\mathbf{e}$ is the unit vector of exciton polarization, $\mathbf{k}_{\|}$ and $\mathbf{k_{\rm exc}}$ are in-plane wavevectors of light and exciton, respectively, and $\mathbf{E}^{\rm up}_{n,\mathbf{k}_{\|}}$ is the electric field of $n$-th photonic mode at the upper surface of the PCS. Importantly, exciton-photon interaction is allowed only under condition $\mathbf{k_{\rm exc}}=\mathbf{k}_{\|}+\mathbf{G}$, where $\mathbf{G}$ is the reciprocal lattice vector. Finally, we formulate the total Hamiltonian and proceed with its diagonalization applying the generalized Hopfield transformation~\cite{hopfield1958theory} which can be reduced to a non-Hermitian eigenvalue equation.

We apply the procedure for characterization of band structure and damping rates of exciton-polaritons being formed in the vicinity of the BIC. At low temperatures of about $4$~K energy of A-type bright exciton in $\rm WSe_2$ is of order of  $E_{\rm exc}=1.74$~eV~\cite{wang2015spin} and its dispersion can be neglected at the scales of the problem. The radiative $\tau_{\rm exc,R}$ and non-radiative $\tau_{\rm exc,NR}$ lifetimes  can estimated as $1$~ps~\cite{palummo2015exciton} and $1$~ns~\cite{robert2016exciton}. For a bare exciton, radiative channel dominates and leads to the damping rate $\hbar/(2\tau_{\rm exc}) = 0.33$~meV. However, when the TMDC is strongly coupled to the PCS, the radiative channel of exciton into the photonic system is enabled and it dominates with respect to direct radiation into free space~\cite{poddubny2011spontaneous}. It leads to renormalization of exciton radiative lifetime of exciton used as a parameter for eigenmode procedure. Therefore, in calculations we use non-radiative lifetime for the exciton damping rate $\hbar/(2\tau_{\rm exc, NR})= 0.33\ \mu$~eV. Oscillator strength depends on $\tau_{\rm exc,R}^{-1}$ and is about $6.5\times10^{-12}\ {\rm cm\cdot eV}^2$ (see Supplemental materials for more details). We focus on the off-$\Gamma$ BIC marked with a black square in Fig.~\ref{fig:2} which has TE polarization which leads to better coupling with in-plane excitons according to Eq.~\ref{eq:1}. We choose an off-$\Gamma$ BIC because both its frequency and wavenumber can be effectively tuned by reshaping the structure geometry.  We tune the BIC frequency to a resonance with $E_{\rm exc}$ adjusting the height of the PCS to the value of $H=418$ nm. For numerical simulations we limit the basis to the in-plane linearly polarized exciton and the set of photonic modes used for calculation of Fig.~\ref{fig:2}.

The spectrum of energies $\hbar\omega$ and inverse lifetimes $\hbar/(2\tau)$ of upper and lower exciton-polariton branches (UP and LP, respectively) calculated by means of the GME is shown in Fig.~\ref{fig:3}(a,b). Figure~\ref{fig:3}(a) demonstrates strong coupling between the exciton and the off-$\Gamma$ BIC which manifests itself as an avoided resonance crossing with Rabi splitting of order of {$3$ meV}. The radiation losses of both exciton-polariton branches are shown in Fig.~\ref{fig:3}(b) in comparison with the losses of photonic and exciton modes. As it can be seen, for specific values of $k_x$ the lifetime of polariton modes can exceed the bare exciton lifetime by almost three orders of magnitude and reaches $0.66$ ns. Such giant enhancement is the special effect intrinsic to optical bound states in the continuum. As it was discussed before, direct radiation of excitons into free space is strongly suppressed due to coupling to the PCS. At the same time BICs are entirely uncoupled to the radiation continuum. In total, this leads to suppression of all possible mechanisms of radiation resulting in giant LP lifetime comparable with the non-radiative lifetime of a bare exciton.

\begin{figure} [t]
\begin{minipage}{0.99\linewidth}
\begin{center}
\includegraphics[width=0.95\linewidth]{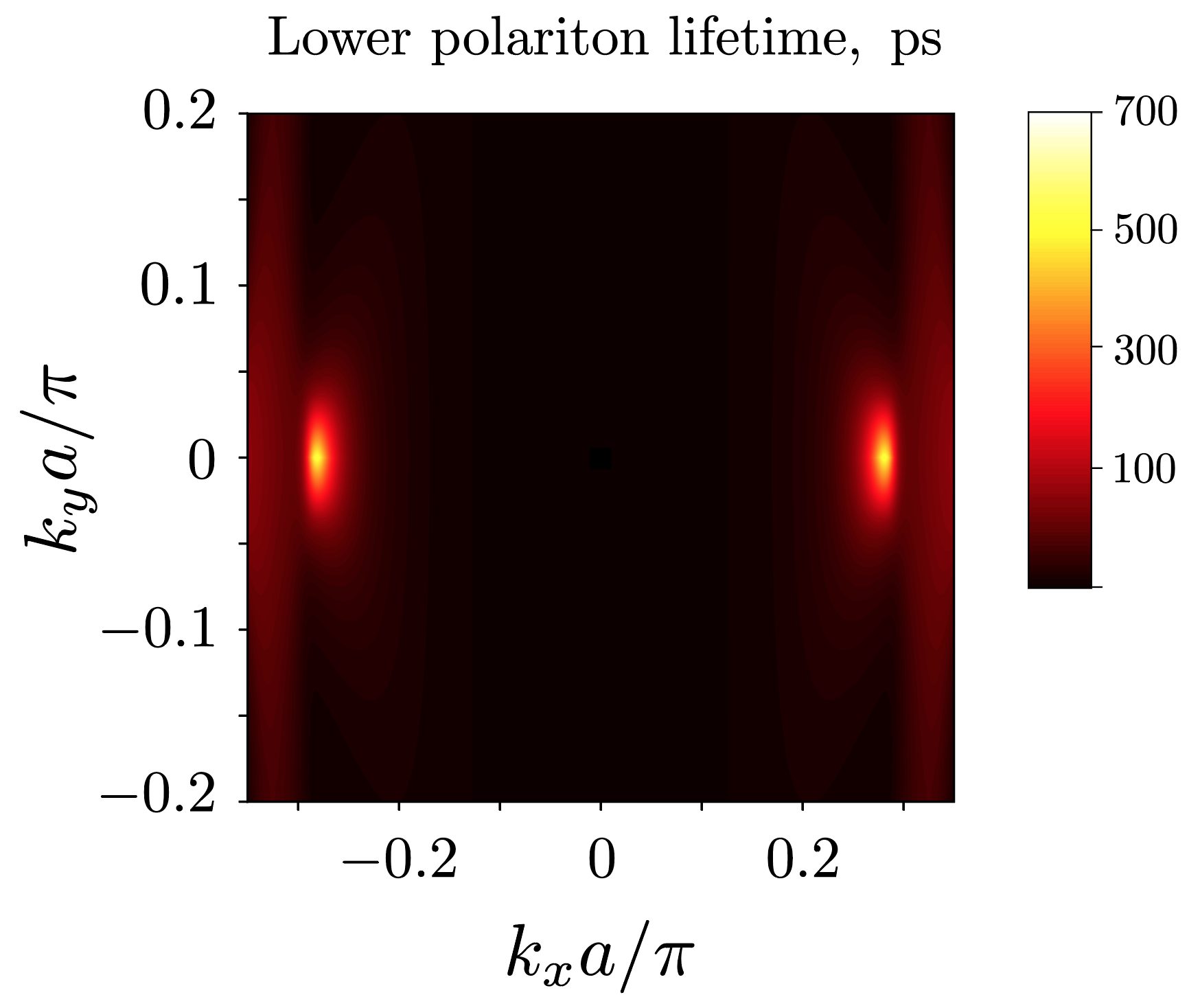}
\end{center}
\caption{Map of dependence of lifetime of the lower polariton mode (red curve in Fig.\ref{fig:3}) on in-plane wavevector $\mathbf{k}$. Calculations are performed by means of the guided-mode expansion method.
}
\label{fig:4}
\end{minipage}
\end{figure}

The most important, Fig.~\ref{fig:3}(b) shows the maximal lifetime can be realized not at the center or the edge of the Brillouin zone, but at the point of phase space, where the group velocity of the mode is finite. For the LP branch this practically means that polaritons can condense to a long-lived state with nonzero energy flow at low temperatures, resulting in moving condensate. While, usually moving condensate is realized by applying the resonant optical pump with fixed in-plane wavevector, in the case of off-$\Gamma$ BIC, the condensation to a propagating state can be achieved with non-resonant pump (either optical or electrical). 

Experimental methods of observation of strong light-matter coupling include angle-resolved reflection spectroscopy. We model such experiment applying the Fourier Modal Method \cite{li1996} with adaptive spatial resolution \cite{vallius2002} which is based on expansion of electromagnetic field and the permittivity function into the Fourier series. Reflectance is obtained via solving wave equation in the discretized Fourier space with proper boundary conditions. Increased efficiency enabled by using the adaptive coordinates  allows to achieve $10^{-4}$ relative error  taking into account $10$ Fourier harmonics in the expansion. To obtain peak positions and their linewidths we fit the reflectance spectrum to Fano lineshape using Levenberg-Marquardt algorithm~\cite{rybin2009fano}. More details can be found in the Supplemental materials.

The real and imaginary part of eigenfrequency of UP and LP branches extracted by fitting procedure are shown in Fig.~\ref{fig:3}(c,d). We see that the GME method and the reflectance spectrum calculations show good agreement providing the same value of Rabi splitting and the 3-fold enhancement of LP lifetime with respect to a bare exciton. However, the maximal values of the damping rate extracted by the fitting method are one order smaller than those for the eigenmode analysis. The difference is the result of approximations used for the GME calculations. First, the GME approach does not consider the leaky modes of the PCS as basis modes leading to  numerical errors of order of $10^{-6}$ eV which is relevant in our case. Second, the quantum formalism used for calculations of light-matter coupling does not take into account the intrinsic photonic modes of the TMDC.

We derive an analytical estimate for the maximal lifetime of the LP mode by using a simple two-level toy model which considers only one photonic and one exciton mode. While the Rabi splitting $|V|$ (see Eq.~\ref{eq:1}) is smaller than detuning $\Delta=\hbar\omega_{\rm ph} - \hbar\omega_{\rm exc}$, mixing of photon and exciton inverse lifetimes is determined by the ratio $(|V|/\Delta)^2$. Interplay between 
the radip increase of the photonic mode lifetime in the vicinity of the BIC and enhancement of coupling to the exciton via decrease of the detuning parameter results in formation of an extremal point in the $\mathbf{k}$-space where the LP lifetime is maximal. Simple estimations (see Supplemental materials for more details) show that LP lifetime is limited 
\begin{equation}
{\tau_{\rm LP}} \le  \xi |V|^{-1} (\tau_{\rm exc,NR})^{{1}/{2}}  ,
\label{eq:2}
\end{equation}
where  $\xi\sim 1.66\cdot 10^{-7}$~eV$\cdot$ s$^{1/2}$  is the constant determined by the PCS design. Equation~\ref{eq:2} provides the estimate ${\tau_{\rm LP}}\le 1.8$~ns, which is in a good agreement with values $0.7$~ns and $3$~ns obtained by means of the GME and reflection calculations, respectively.  

Finally, we calculate the dependence of lower-polariton lifetime on in-plane wavevector $\mathbf{k}={k_x,k_y}$ in the 2D Brillouin zone by means of the eigenmode analysis (see Fig.~\ref{fig:4}). One can see that lifetime value exhibits maximum at $k_x \simeq \pm 0.29\pi/a, k_y = 0$ and decreases smoothly in the vicinity of these points. Importantly, geometry of one-dimensional PCS allows to realize a BIC only on $k_y=0$ or $k_x=0$ lines of phase space. To enable full control of BIC position within the phase space of wave vectors wide class of two-dimensional periodic photonic structures can be used~\cite{zhen2014topological}. Therefore, formation of long-living polariton states can be achieved for any value of $\mathbf{k}$ which leads to absolute freedom in engineering of polariton condensates with desired value and direction of energy flow and even negative group velocities can be achieved.

In conclusion, we have proposed an experimentally feasible scheme to achieve strong coupled exciton-photon system in a two-dimensional nanostructure comprising a TMDC monolayer and a periodic photonic nanostructure. Importantly, this scheme does not require the growth of Bragg mirrors, which substantially simplifies the fabrication. Moreover, we have shown that it also allows the polariton condensation at the finite momenta, which opens possibilities for the non-resonant excitation of moving polariton condensates. We believe, these findings open new avenues for the applications of strong light-matter coupling at the nanoscale.

\begin{acknowledgments}
Theoretical studies have been supported by the megagrant (14.Y26.31.0015) and GosZadanie (3.1365.2017/4.6). Simulations of scattering spectrum have been supported by the Russian Science Foundation (17-12-01581).
\end{acknowledgments}

\bibliography{ExcBicBib}

\newpage
\appendix

\section{SUPPLEMENTAL MATERIALS}

In the Supplemental materials we (i) derive dependence of oscillator strength on exciton radiation lifetime, (ii) apply two-level system approximation to estimate maximal lifetime of exciton-polariton mode and (iii) calculate surface conductivity of TMDC and use it to analyse reflectance spectrum of TMDC coupled to optical bound states in the continuum.

\section{Oscillator strength}

In this section we derive the expression for oscillator strength describing optical transitions for excitons in low-dimensional ultra-thin quantum structures.

Using Fermi's golden rule and neglecting exciton dispersion we calculate the inverse radiation lifetime of exciton as~\cite{ivchenko2005optical}

\begin{equation}
\gamma_{\rm exc,R}=\frac{2\pi \omega_{\rm exc}}{\hbar c}\left( \frac{e|p_{cv}|}{m\omega_{\rm exc}} \right)^2\left[\int dz\ \Phi(z) \cos(\omega_{\rm exc} z/c)\right]^2,
\end{equation}
where $\omega_{\rm exc}$ is the exciton frequency, $m$ is the mass of the exciton, $\Phi(z)$ is smooth envelope of excitonic wavefunction and $|p_{cv}|$ is matrix element of momentum operator for optical transitions between the conduction and the valence band. Notably, $\cos(\omega_{\rm exc} z/c)$ can be replaced by unity because of small structure thickness.

The oscillator strength of excitonic transition per unit area $F/S$ is generally defined as~\cite{andreani2007light} 
\begin{equation}
\frac{F}{S}= \frac{2}{m\hbar \omega_{\rm exc}}|p_{cv}|^2\left[\int dz\ \Phi(z) \right]^2.
\end{equation}

We introduce the definition of renormalized oscillator strength $f$ which is more convenient for our problem
\begin{equation}
f= \frac{\pi\hbar^2e^2}{m}\frac{F}{S}.
\end{equation}

Therefore, the oscillator strength depends  on the exciton radiative lifetime as
\begin{equation}
f= c \hbar^2 \gamma_{\rm exc,R}.
\end{equation}

\section{Two-level analytical approximation}

In this section we calculate the maximal lifetime of the lower polariton branch, which dispersion is shown in Fig.~3 of the main text. We investigate a toy two-level model considering the interaction between one photonic and one exciton mode. Complex eigenenergies of UP and LP which represent hybridized light-matter states are
\begin{multline}
\hbar\Omega_{\rm UP,\ LP} = \frac{\hbar\omega_{\rm ph}+\hbar\omega_{\rm exc}}{2}  - i\frac{\hbar\gamma_{\rm ph}+\hbar\gamma_{\rm exc}}{2} \mp \\  \left( {|V|^2} + \left[\frac{\hbar\omega_{\rm ph}-\hbar\omega_{\rm exc}}{2} -i \frac{\hbar\gamma_{\rm ph}-\hbar\gamma_{\rm exc}}{2} \right]^2 \right)^{1/2},
\end{multline}
where $\omega_{\rm ph, exc}$ and $\gamma_{\rm ph,exc}$ are frequencies and inverse lifetimes of bare photonic and exciton modes, respectively, and $|V|$ is Rabi splitting, determined by Eq.~1 of the main text. Here, we use $\gamma_{\rm exc}=\gamma_{\rm exc,NR}$ because of suppression of direct radiation of exciton to free space. 

We focus on the lower polariton mode, but the derivation is valid for the upper polariton as well. In the weak coupling regime, when LP dispersion is close to the bare photonic mode, the damping rate is slightly perturbated due to mixing with the exciton  
\begin{equation}
\hbar\gamma_{\rm LP} = \hbar\gamma_{\rm ph} \left(1 - \frac{|V|^2}{\Delta^2}\right) + \hbar\gamma_{\rm exc,NR} \frac{|V|^2}{\Delta^2},
\end{equation}
where $\Delta = \hbar\omega_{\rm ph}- \hbar\omega_{\rm exc}$ is detuning. 

We analyze narrow band of wave vectors close to the BIC position where the condition $\gamma_{\rm ph} \ll \gamma_{\rm exc,NR}$ is fulfilled. The maximal lifetime is given by the equation
\begin{equation}
\frac{d\gamma_{\rm ph}}{dk_x} = 2 \gamma_{\rm exc,NR}\frac{|V|^2}{\Delta^3} \frac{d \Delta}{dk_x},
\end{equation}
where we use the fact that $|V|$ depends on wavevector smoothly.
The detuning in a vicinity of the wavevector of the BIC $k_{\rm BIC}$ is linear $\Delta = \hbar v_{\rm g} (k_x-k_{\rm BIC})$ (see Fig.3 of the main text), while the singularity of $\gamma_{\rm ph}$ is quadratic~\cite{Yuan_2017_OB_in_BIC,bulgakov2017topological}  $\gamma_{\rm ph} = \alpha (k_x-k_{\rm BIC})^2$. Constant $\alpha$ and group velocity $v_{\rm g}$ depend only on the design of the PCS and can be calculated numerically.
Finally, the maximal lifetime of the LP mode is realized for the wavevector $k_x^*$
\begin{equation} 
k_x^* = k_{\rm BIC} - \left(\frac{\gamma_{\rm exc,NR} |V|^2}{\alpha(\hbar v_{\rm g})^2}\right)^{1/4},
\end{equation}
and the minimal value of $\gamma_{\rm LP}$ is 
\begin{equation} 
\gamma_{\rm LP}(k_x^*)=2|V|\frac{\sqrt{\alpha}}{\hbar v_{\rm g}} (\gamma_{\rm exc,NR})^{1/2},
\end{equation}
which implies the constant $\xi$ defined in the main text is equal to $(\hbar v_{\rm g})/(8\alpha)^{1/2}$.

\section{Reflectance spectrum}

\begin{figure} [t]
\begin{minipage}{0.99\linewidth}
\begin{center}
\includegraphics[width=0.95\linewidth]{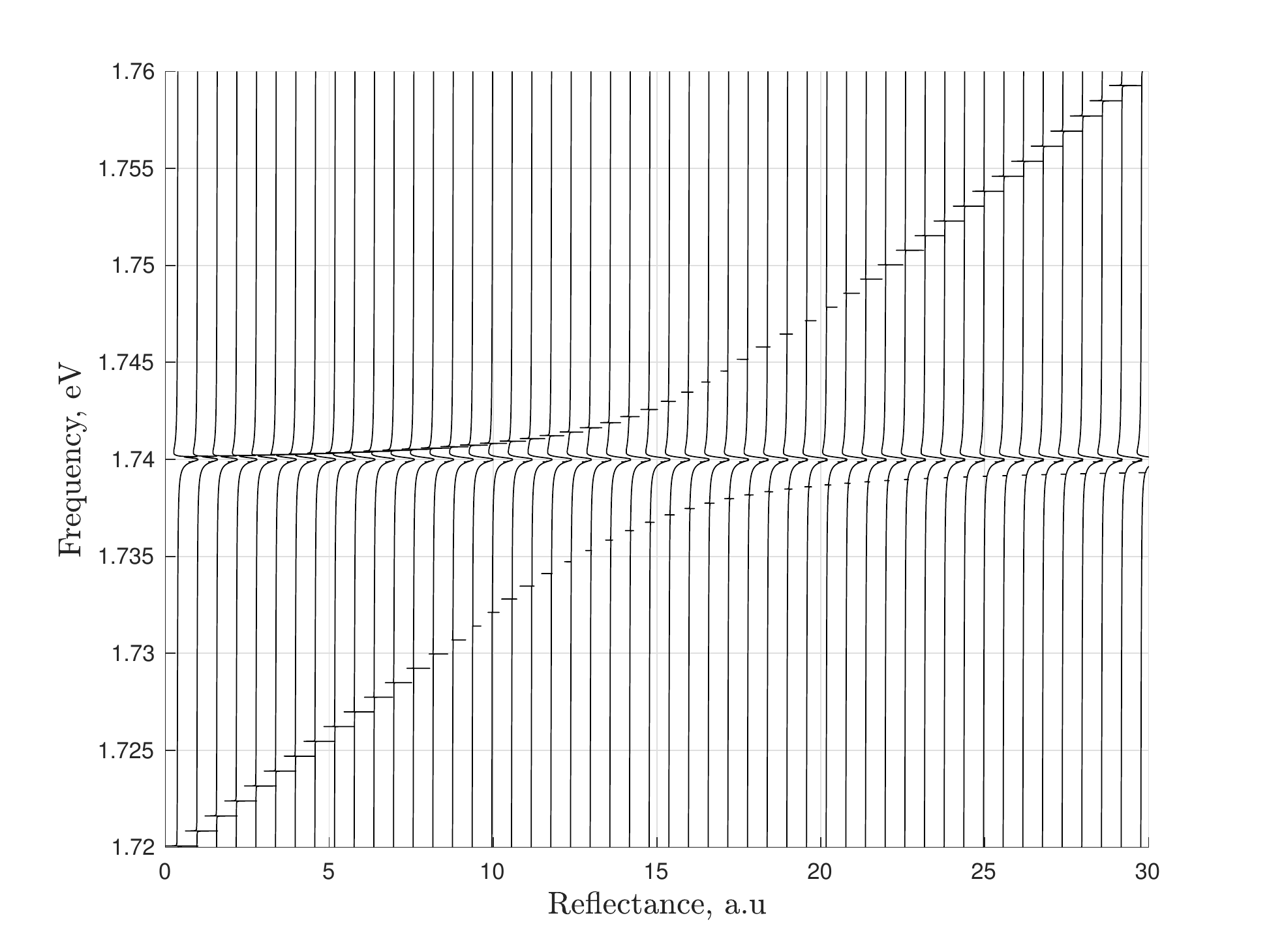}
\end{center}
\caption{ Reflectance spectrum of $\rm WSe_2$ TMDC placed on top of one-dimensional $\rm Ta_2O_5$ PCS. Each spectrum is shifted in horizontal direction by 0.6. } 
\label{fig:1}
\end{minipage}
\end{figure}

In this section we derive the expression for surface conductivity of TMDC monolayer that can be used for calculations of reflectance spectrum. 

Reflection coefficient of TE-polarized light $r_{\rm TE}$ for 2D monolayers which optical properties are governed by excitonic response is~\cite{ivchenko2005optical}
\begin{equation}
r_{\rm TE} = \frac{i\gamma_{\rm exc, R}^{\rm TE}}{\omega_{\rm exc}-\omega-i(\gamma_{\rm exc, R}^{\rm TE}+\gamma_{\rm exc, NR})}.
\end{equation}
Here modified inverse radiation lifetime for TE-polarized light depends on angle of incidence $\theta$ as
\begin{equation}
\gamma_{\rm exc, R}^{\rm TE} = \frac{\gamma_{\rm exc, R}}{\cos{\theta}}.
\end{equation}

Surface conductivity of 2D material can be expressed via the reflection coefficient using Maxwell's boundary conditions for magnetic field components~\cite{merano2016fresnel} 
\begin{equation}
\sigma_{\rm TMDC}=\frac{c}{4\pi}\left[1+\frac{r_{\rm TE}-1}{1+r_{\rm TE}}\right]\cos{\theta},
\end{equation}
where it was assumed that the TMDC monolayer is surrounded by air.

Combining all equations we arrive at the formula for surface conductivity of TMDC layer
\begin{equation}
\sigma_{\rm TMDC} = \frac{ic}{2\pi}\frac{\gamma_{\rm exc, R}}{\omega_{\rm exc}-\omega-i\gamma_{\rm exc, NR}}.
\label{eq:1}
\end{equation}

Reflectance spectrum for TMDC-PCS structure with parameters given in the main text calculated by means of Eq.~\ref{eq:1} is shown in Fig.~\ref{fig:1}.

\end{document}